\documentclass{elsart}

 \usepackage{graphicx}

\usepackage{amssymb}

\begin{document}

\begin{frontmatter}



\title{Density fluctuations on mm and Mpc scales}


\author{N. P. Basse \corauthref{cor1}}
\ead{basse@psfc.mit.edu}
\ead[url]{http://www.psfc.mit.edu/people/basse/}
\corauth[cor1]{Corresponding author. Tel.: +1 617 253 5523; fax:
+1 617 253 0627.}

\address{Plasma Science and Fusion Center, Massachusetts Institute of Technology, Cambridge, MA 02139, USA}

\begin{abstract}
We will in this paper report on suggestive similarities between
density fluctuation power versus wavenumber on small (mm) and
large (Mpc) scales.

The small scale measurements were made in fusion plasmas and
compared to predictions from classical fluid turbulence theory.
The data is consistent with the dissipative range of 2D
turbulence. Alternatively, the results can be fitted to a
functional form that can not be explained by turbulence theory.

The large scale measurements were part of the Sloan Digital Sky
Survey galaxy redshift examination. We found that the equations
describing fusion plasmas also hold for the galaxy data.

The comparable dependency of density fluctuation power on
wavenumber in fusion plasmas and galaxies might indicate a common
origin of these fluctuations.
\end{abstract}

\end{frontmatter}

\section{Introduction}
\label{sec:intro}

If one were to make a survey of where we are, what we know and
what we do not know about magnetically confined fusion plasmas,
turbulence would certainly be an area marked 'Here Be Monsters'.
Cross-field transport (perpendicular to the main magnetic field)
assuming that only binary particle collisions contribute is called
neoclassical transport \cite{hinton}. This transport level
includes effects associated with toroidal geometry. However, in
general the measured transport is several orders of magnitude
larger than the neoclassical one, especially for electrons. This
phenomenon has been dubbed anomalous transport and is subject to
intense studies on most experimental fusion devices
\cite{wootton}. Anomalous transport is believed to be driven by
turbulence in the plasma.

It is generally thought that turbulence creates fluctuations
visible in most plasma parameters. Therefore a concerted effort
has been devoted to the study of fluctuations and their relation
to the global (and local) plasma confinement quality.

In this paper we study electron density fluctuation power versus
wavenumber (also known as the wavenumber spectrum) in the
Wendelstein 7-AS (W7-AS) stellarator \cite{renner}. Wavenumber
spectra characterize the nonlinear interaction between turbulent
modes. The fluctuations were measured using the localized
turbulence scattering (LOTUS) diagnostic \cite{saffman,basse1}.

As we shall see, the density fluctuation power $P$ decreases
exponentially with increasing wavenumber $k$ on mm scales in
fusion plasmas

\begin{equation}
P(k) \propto \frac{1}{k} \times e^{-nk}, \label{eq:exp_decay}
\end{equation}

where $n > 0$ is a constant having a dimension of length and $k =
2\pi/\lambda$, where $\lambda$ is the corresponding wavelength.
This was initially noted using the simplified form

\begin{equation}
P(k) \propto e^{-nk} \label{eq:exp_decay_simpl}
\end{equation}

in Ref. \cite{basse2}. Eq. (\ref{eq:exp_decay_simpl}) also holds
for density fluctuations in the Tore Supra tokamak
\cite{hennequin}.

Having the exponential structure of the wavenumber spectrum in
mind, we were intrigued to see a figure in KVANT \cite{kvant}, a
magazine published by the Danish Physical Society, showing the
root-mean-square density fluctuation amplitude of galaxies from
the Sloan Digital Sky Survey (SDSS) \cite{sdss1} versus distance
that seemed to display the same exponential behavior as we found
in fusion plasmas. This remarkable similarity prompted us to apply
the same analysis to the SDSS data as we had previously used for
our fusion plasma measurements.

The paper is organized as follows: In Section \ref{sec:fusion} we
summarize our wavenumber spectrum measurements in fusion plasmas.
Thereafter we describe inflation and the SDSS wavenumber spectrum
in Section \ref{sec:galax}. We discuss the results in Section
\ref{sec:disc} and finally state our conclusions in Section
\ref{sec:conc}.

\section{Fusion plasmas}
\label{sec:fusion}

\begin{center}
\begin{figure}
\includegraphics[width=12cm]{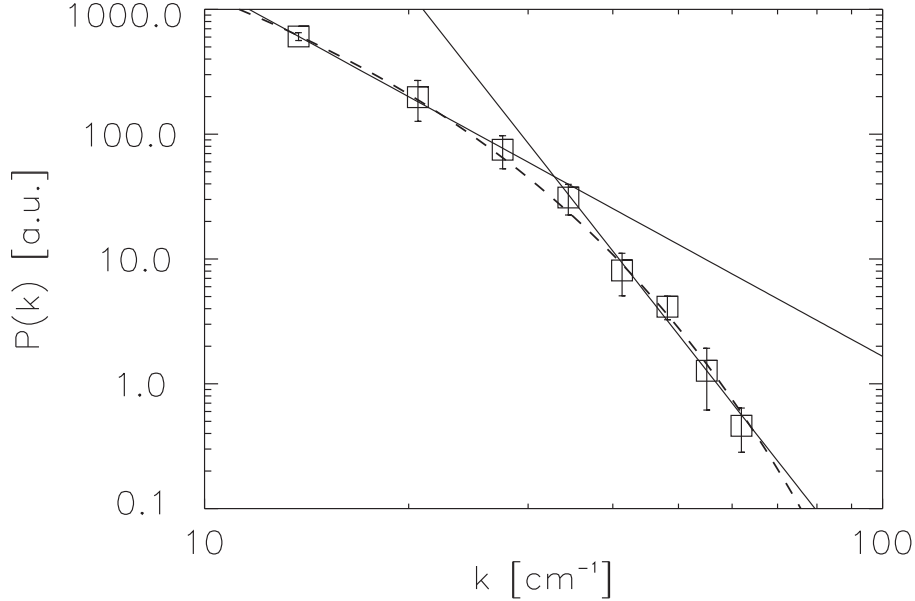}
\caption{\label{fig:w7as} Wavenumber spectrum of broadband
turbulence in W7-AS measured using the LOTUS diagnostic. Squares
are measured points. Solid lines are power-law fits to the three
smallest and five largest wavenumbers, the dashed line is a fit to
Eq. (\ref{eq:exp_decay}). The power-law fit grouping of points is
the only one where convergence is obtained.}
\end{figure}
\end{center}

A wavenumber spectrum of turbulence in W7-AS is shown in Fig.
\ref{fig:w7as}. The measured points are shown along with two
power-law fits

\begin{equation}
P(k) \propto k^{-m}, \label{eq:pow_decay}
\end{equation}

where $m$ is a dimensionless constant. The power-law fits are
shown as solid lines and an exponential fit to Eq.
(\ref{eq:exp_decay}) is shown as a dashed line. The power-law fits
are motivated by classical fluid turbulence theory where one
expects wavenumber spectra to exhibit power-law behavior with
exponents $m$ depending on the dimension of the observed
turbulence:

\begin{itemize}
\item 3D: Energy is injected at a large scale and redistributed
(cascaded) by nonlinear interactions down to a small dissipative
scale. In this case, the energy spectrum in the inertial range
$E(k) \propto k^{-5/3}$ \cite{frisch}. \item 2D: Here, two
power-laws exist on either side of the energy injection scale. For
smaller wavenumbers, the inverse energy cascade obeys $E(k)
\propto k^{-5/3}$ and for larger wavenumbers, the enstrophy
cascade follows $E(k) \propto k^{-3}$ \cite{antar1}. \item 1D:
Energy is injected at a large scale and dissipated at a small
scale; $E(k) \propto k^{-2}$ \cite{neumann}.
\end{itemize}

Our measured power is equivalent to the $d$-dimensional energy
spectrum $F_d(k)$ \cite{tennekes,frisch,antar2}

\begin{eqnarray}
P(k) = F_d(k) = \frac{E(k)}{A_d} \nonumber \\ \nonumber \\ A_1 = 2
\hspace{2cm} A_2 = 2\pi k \hspace{2cm} A_3 = 4\pi k^2,
\label{eq:e_spec}
\end{eqnarray}

where $A_d$ is the surface area of a sphere having radius $k$ and
dimension $d$. Usually on would assume that $d = 2$ in fusion
plasmas, since transport along magnetic field lines is nearly
instantaneous. The fits to Eq. (\ref{eq:pow_decay}) in Fig.
\ref{fig:w7as} yield exponents $m$ = 3 (small wavenumbers) and 7
(large wavenumbers). A similar behavior has previously been
reported in Ref. \cite{honore} where it was speculated that the
wavenumber value at the transition between the two power-laws
should correspond to a characteristic spatial scale in the plasma.
The only length scale close to the transitional value was found to
be the ion Larmor radius $\rho_i$.

The spectrum at small wavenumbers is roughly consistent with the
inverse energy cascade in 2D turbulence, $F_2(k) \propto
k^{-8/3}$. The exponent at large wavenumbers does not fit into
this framework. However, for very large wavenumbers one enters the
dissipation range; here, it has been argued that the energy
spectrum could have one of the following dependencies

\begin{equation}
E_{\rm Neumann}(k) \propto e^{-ak} \hspace{2cm} E_{\rm
Heisenberg}(k) \propto k^{-7}, \label{eq:diss}
\end{equation}

where $a > 0$ is a constant having a dimension of length (see Ref.
\cite{neumann} and references therein). The energy spectrum
proposed by J. von Neumann was what initially inspired us to
investigate an exponential decay of $P(k)$ in Ref. \cite{basse2}.
Fitting all wavenumbers to Eq. (\ref{eq:exp_decay}), $E_{\rm
Neumann}(k)/A_2$, we find that $n$ = 0.1 cm or a wavenumber of 55
cm$^{-1}$. Alternatively, the transitional wavenumber found at the
separation between the two power-laws is 31 cm$^{-1}$ (0.20 cm).
The expression $E_{\rm Heisenberg}(k)/A_2$ yields $m$ = 8, which
is close to the experimental value $m$ = 7 for large wavenumbers.
Calculating the ion Larmor radius at the electron temperature
$\rho_s$ for this case we find that it is 0.1 cm, i.e. the same
order of magnitude as the spatial scales found above. We used
$\rho_s$ instead of $\rho_i$ because ion temperature measurements
were unavailable.

Currently we can think of three possible explanations for the
behavior of the wavenumber spectrum:

\begin{enumerate}
\item We observe 2D turbulence and the transition between the two
power-laws occurs at a spatial scale where the inverse energy
cascade develops into the dissipation range. However, the
enstrophy cascade is not accounted for in this case. \item We
observe 2D turbulence in the dissipation range described by a
single exponential function as proposed by J. von Neumann. \item
Turbulence theory does not apply. The transition between two
power-laws or the characteristic scale found using a single
exponential function (Eq. (\ref{eq:exp_decay_simpl})) indicates
that one scale dominates the turbulent dynamics in the wavenumber
range studied.
\end{enumerate}

\section{Galaxies}
\label{sec:galax}

\subsection{Inflation}
\label{subsec:infla}

Dramatic developments have taken place in cosmology over the last
decade, lending increasing support to the paradigm of inflation as
an explanation for what took place before the events described by
the big bang theory \cite{guth}. Inflation solved the so-called
horizon and flatness problems, but was at odds with earlier
observations indicating that the ratio of the mass density of the
universe to the critical value, the density parameter $\Omega$,
was 0.2-0.3, while inflation predicted it should be 1:

\begin{eqnarray}
\Omega = \frac{\rho}{\rho_c} \nonumber \\ \nonumber \\ \Omega < 1:
{\rm open} \hspace{2cm} \Omega = 1: {\rm flat} \hspace{2cm} \Omega
> 1: {\rm closed},
\label{eq:mass}
\end{eqnarray}

where $\rho_c = 3H_0^2/8\pi G$ is the critical mass density, $H_0
= 70$ km/s/Mpc is the Hubble parameter observed today and $G$ is
I. Newton's gravitational constant \cite{kinney}. However, new
measurements in the late 1990's lead to a drastic modification of
$\Omega$: Observations of type Ia supernovae (SN) showed that the
separation velocity between galaxies was speeding up, not slowing
down as would be expected for an open universe. The underlying
explanation for this accelerated expansion is not understood, but
it seems that the universe contains large quantities of negative
pressure substance, creating a gravitational repulsion driving the
expansion. This negative pressure material is called dark energy,
the total density of dark energy $\Omega_{\Lambda}$ is 0.7. The
existence of dark energy is equivalent to the cosmological
constant $\Lambda$ introduced by A. Einstein. The dark matter
density $\Omega_d$ is 0.25 and the baryonic matter density
$\Omega_b$ is 0.05, so the total density is very close (or equal)
to the critical density. The SN Ia data is supported by
measurements of nonuniformities in the cosmic microwave background
(CMB) radiation. The CMB anisotropy is due to the presence of tiny
primordial density fluctuations at the time of recombination,
where atoms formed. At that point in time the age of the universe
was about 300,000 years and the temperature was 3000 K. The
structures observed in the CMB are called acoustic peaks, and the
simplest versions of inflation all reproduce these structures
quite accurately. The acoustic peaks can not be modelled by
assuming that the universe is open.

\subsection{Wavenumber spectrum}
\label{subsec:wano}

\begin{center}
\begin{figure}
\includegraphics[width=12cm]{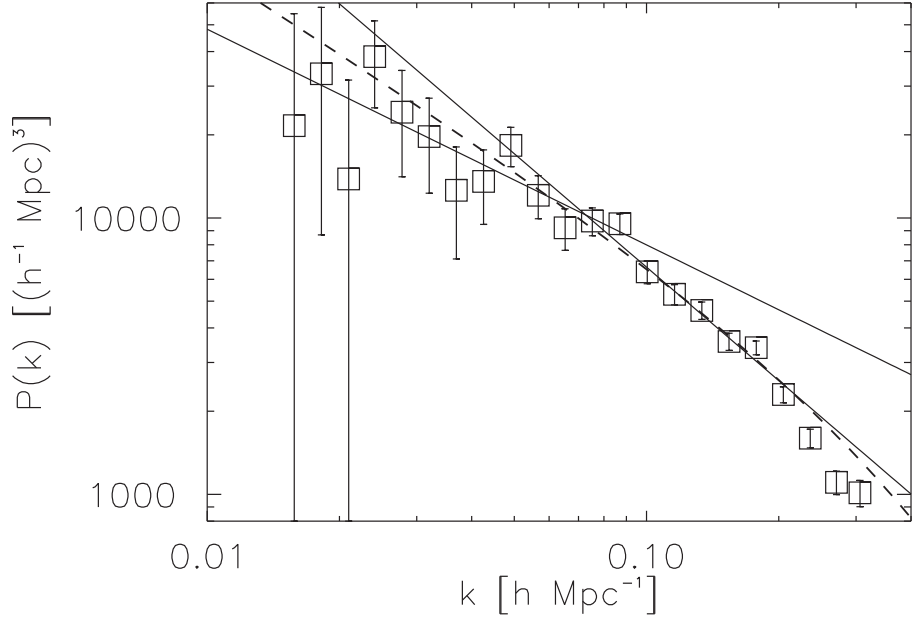}
\caption{\label{fig:sdss} Wavenumber spectrum of galaxies measured
by the SDSS Team. Squares are measured points. Solid lines are
power-law fits, the dashed line is a fit to Eq.
(\ref{eq:exp_decay}). The power-law fit grouping of points is
chosen so the combined, normalized, $\chi^2$ of the fits is
minimized. The data is taken from Ref. \cite{tegmark3}.}
\end{figure}
\end{center}

A study of density fluctuations on large scales using 205,443
galaxies has been published by the SDSS Team in Ref.
\cite{tegmark1}, see Fig. \ref{fig:sdss}. 3D maps of the universe
are provided by the SDSS galaxy redshift survey, observing about a
quarter of the celestial sphere using a 2.5 m telescope and a
charge-coupled device (CCD) camera. The galaxies had a mean
redshift $z \approx$ 0.1, corresponding to light emitted 1-2 Gyr
ago \cite{kinney}. Fixing some cosmological parameters to
Wilkinson Microwave Anisotropy Probe (WMAP) satellite values
\cite{bennett,spergel,verde} one finds - using physics based
models - that the wavenumber spectrum measurements were fitted by
a matter density $\Omega_m = \Omega_d + \Omega_b = 0.295 \pm
0.0323$. In this case $h = H_0$/(100 km/s/Mpc) = 0.72 was assumed
and it was observed that the wavenumber spectrum was not well
characterized by a single power-law.

A follow-up paper by the SDSS Team, Ref. \cite{tegmark2}, combined
non-CMB measurements (SDSS) with CMB measurements (WMAP) to
constrain free parameters of cosmological models and break CMB
degeneracies in parameter space. This resulted in $\Omega_m = 0.30
\pm 0.04$ and $h = 0.70_{-0.03}^{+0.04}$. Adding the SDSS
information more than halved WMAP-only error bars on some
parameters, e.g. the Hubble parameter and matter density.



The data presented in Fig. \ref{fig:sdss} has been taken from M.
Tegmark's homepage \cite{tegmark3}. According to the
recommendation by the SDSS Team \cite{tegmark1}, the three largest
wavenumbers shown are not used in the fits described below.

As we did for the W7-AS data in Section \ref{sec:fusion}, we fit
the SDSS measurements to two power-laws (Eq. (\ref{eq:pow_decay}))
or a single exponential function (Eq. (\ref{eq:exp_decay})). The
power-law fits are shown as solid lines, the exponential fit as a
dashed line.

The power-law fits yield exponents $m$ = 0.8 (small wavenumbers)
and 1.4 (large wavenumbers). The wavenumber ranges were determined
by minimizing the combined, normalized, $\chi^2$ of the fits. As
the SDSS Team found, a single power-law can not describe the
observations. The exponents are not close to the ones governing
fluid turbulence discussed in Section \ref{sec:fusion}. The
transitional wavenumber is 0.09 h Mpc$^{-1}$, corresponding to a
length of 67 h$^{-1}$ Mpc.

We find the characteristic length from an exponential fit to be
$n$ = 2 h$^{-1}$ Mpc or a wavenumber of 3 h Mpc$^{-1}$.

\section{Discussion}
\label{sec:disc}

The fact that density fluctuations on small (fusion plasmas) and
large (galaxies) scales can be described by an exponential
function might indicate that plasma turbulence at early times has
been expanded to cosmological proportions. A natural consequence
of that thought would be to investigate fluctuations in
quark-gluon plasmas (QGPs) corresponding to even earlier times.
However, experimental techniques to do this are not sufficiently
developed at the moment due to the extreme nature of QGPs.

It is fascinating that wavenumber spectra over wider scales peak
at small wavenumbers and decrease both above and below that peak.
This is seen both in fusion plasmas \cite{basse3} and for
galaxies, see e.g. Fig. 38 in Ref. \cite{tegmark1}. Turbulence
theory in 1D or 3D would interpret the peak position as the scale
where energy is injected.

Fitting wavenumber spectra to power-laws is based on fluid
turbulence theories, but in general care must be taken when
interpreting the outcome: We know that an exponential function can
be Taylor expanded to an infinite power series:

\begin{equation}
P(k) \propto e^{-nk} = \sum_{i=0}^{\infty} \frac{(-nk)^i}{i!}.
\label{eq:taylor}
\end{equation}

So locally, i.e. for a small range of wavenumbers, an exponential
dependency can be masked as a power-law; the exponent would vary
as a function of the wavenumber range selected.

We favor the exponential functions over power-laws as descriptors
of the data, either including an algebraic prefactor as in Eq.
(\ref{eq:exp_decay}) or using a pure exponential function as in
Eq. (\ref{eq:exp_decay_simpl}): In the latter case, fits yield $n$
= 0.2 cm for fusion plasmas and 11 h$^{-1}$ Mpc for galaxies. The
exponential decrease of density fluctuation power versus
wavenumber implies that either (2) or (3) in Section
\ref{sec:fusion} could explain the data. Perhaps forcing occurs at
a large scale and transitions either directly to dissipation (2)
or to some effect governed by the scale determined from the
exponential function (3) at smaller scales. The fact that we
obtain $n \simeq \rho_s$ using Eq. (\ref{eq:exp_decay}) for fusion
plasmas supports this conjecture.


\section{Conclusions}
\label{sec:conc}

We have in this paper reported on suggestive similarities between
density fluctuation power versus wavenumber on small (mm) and
large (Mpc) scales.

The small scale measurements were made in fusion plasmas and
compared to predictions from turbulence theory. The data fit Eq.
(\ref{eq:exp_decay}), which is consistent with the dissipative
range of 2D turbulence. Alternatively, the results fit Eq.
(\ref{eq:exp_decay_simpl}) which has a functional form that can
not be explained by turbulence theory.

The large scale measurements were part of the SDSS galaxy redshift
survey. As is the case for fusion plasmas, the galaxy data can be
described by Eqs. (\ref{eq:exp_decay}) and
(\ref{eq:exp_decay_simpl}). The similar dependency of density
fluctuation power on wavenumber might indicate a common origin of
these fluctuations, perhaps from fluctuations in QGPs at early
stages in the formation of the universe.

The cross-disciplinary work presented here is hopefully just the
beginning of an interesting path that can benefit both fields. As
a first step, we will expand our studies to encompass a wider
range of scales, both for fusion plasma and galaxy measurements.

\begin{ack}
This work was supported at MIT by the Department of Energy,
Cooperative Grant No. DE-FC02-99ER54512.
\end{ack}



\end{document}